\documentclass[manuscript]{aastex62}

\usepackage{mhchem}
\usepackage{natbib}
\usepackage[utf8]{inputenc}
\usepackage{placeins}
\usepackage{enumitem}

\graphicspath{{./}{figures/}}

\received{January 1, 2018}
\revised{January 7, 2018}
\accepted{\today}
\submitjournal{ApJ}

\shorttitle{On Simulating Experiments Using Astrochemical Models}
\shortauthors{Shingledecker et al.}

\begin{document}
\title{On Simulating the Proton-Irradiation of O$_2$ and H$_2$O Ices Using Astrochemical-type Models, \\ with Implications for Bulk Reactivity}

\correspondingauthor{Christopher N. Shingledecker}
\email{cns@mpe.mpg.de}

\author[0000-0002-5171-7568]{Christopher N. Shingledecker}
\affil{Center for Astrochemical Studies 
Max Planck Intitute for Extraterrestrial Physics 
Garching, Germany}
\affil{Institute for Theoretical Chemistry 
University of Stuttgart 
Pfaffenwaldring 55, 70569 
Stuttgart, Germany}

\author{Anton Vasyunin}
\affil{Ural Federal University,  Ekaterinburg, Russia}
\affil{Visiting Leading Researcher, Engineering Research Institute 'Ventspils International Radio
Astronomy Centre' of Ventspils University of Applied Sciences, 
 In\v{z}enieru 101, Ventspils LV-3601, Latvia}

\author{Eric Herbst}
\affil{Department of Chemistry 
University of Virginia 
Charlottesville, VA, USA}
\affil{Department of Astronomy 
University of Virginia 
Charlottesville, VA, USA}

\author{Paola Caselli}
\affil{Center for Astrochemical Studies 
Max Planck Intitute for Extraterrestrial Physics 
Garching, Germany}

\begin{abstract}

Many astrochemical models today explicitly consider the species that comprise the bulk of interstellar dust grain ice-mantles separately from those in the top few monolayers. Bombardment of these ices by ionizing radiation - whether in the form of cosmic rays, stellar winds, or radionuclide emission - represents an astrochemically viable means of driving a rich chemistry even in the bulk of the ice-mantle, now supported by a large body of work in laboratory astrophysics.
In this study, using an existing rate equation-based astrochemical code modified to include a method of considering radiation chemistry recently developed by us, we attempted to simulate two such studies in which (a) pure \ce{O2} ice at 5 K and, (b) pure \ce{H2O} ice at 16 K and 77 K, were bombarded by keV \ce{H+} ions.  

Our aims are twofold: (1) to test the capability of our newly developed method to replicate the results of ice-irradiation experiments, and (2) to determine in such a well-constrained system how bulk chemistry is best handled using the same gas-grain codes that are used to model the interstellar medium (ISM). 
We find that our modified astrochemical model is able to reproduce both the abundance of \ce{O3} in the 5 K pure \ce{O2} ice, as well as both the abundance of \ce{H2O2} in the 16 K water ice and the previously noted decrease of hydrogen peroxide at higher temperatures. However, these results require the assumption that radicals and other reactive species produced via radiolysis react quickly and non-diffusively with neighbors in the ice.

\end{abstract}

\section{Introduction} \label{sec:intro}

The upcoming launch of the James Webb Space Telescope (JWST) promises to usher in a new astrochemical ``ice age'' by allowing for the collection of an unprecedented amount of data regarding species frozen onto dust grains in the interstellar medium (ISM). Astrochemical models will surely figure prominently in both interpreting these data as well as in informing future observing proposals. This opportunity, though, calls for models that are equal to the task and, unfortunately, there remain significant uncertainties regarding the chemistry of dust grains and dust grain ice-mantles \citep{cuppen_grain_2017} - a situation that has given it the dubious reputation of ``the last refuge of the scoundrel'' \citep{charnley_molecular_1992}.  

Currently, it is a common practice in modern astrochemical codes to distinguish between species in the top few monolayers of dust grain ice-mantles (the selvedge), and those underneath in the interior (the bulk), though there exists considerable disagreement as to whether, or to what degree, such bulk species are chemically active \citep{cuppen_grain_2017}. The hypothesis that chemistry within the ice-mantle is less significant than in the selvedge rests on the following three main lines of reasoning:

\begin{enumerate}[label=(\alph*)]
\item  species in the bulk are more tightly bound,
\item  a richer chemistry is possible in the selvedge, given the constant exchange of species with the surrounding gas, and
\item  photo-processes, which further drive chemical reactions, are more efficient in the topmost monolayers. 
\end{enumerate}

\noindent
An implicit assumption behind (a) is that reactions in the bulk occur largely via thermal diffusion, as on surfaces, and thus that reaction rates will be low or negligible for species with high binding energies in cold environments. However, some previous experimental studies strongly suggest that reactions in the bulk can quickly occur among neighboring species, without the need of diffusion \citep{baragiola_solid-state_1999,bennett_laboratory_2005,abplanalp_study_2016,ghesquiere_reactivity_2018}. 

Similarly, just as selvedge chemistry is stimulated via photo-processes and interactions with the gas - as noted in (b) and (c) - so too have experiments shown that bombardment by energetic ions, which penetrate solids much more efficiently than photons \citep{gerakines_energetic_2001,spinks_introduction_1990}, causes physicochemical changes that drive a rich chemistry within the ice at low temperatures. This \textit{radiation chemistry} is quite relevant to the ISM, since ices there are continually bombarded by energetic particles of one kind or another, though typically in the form of cosmic rays \citep{indriolo_cosmic-ray_2013,rothard_modification_2017}, stellar winds \citep{madey_far-out_2002,hudson_radiation_2001}, or radionuclide emission \citep{cleeves_radionuclide_2013}. 

Thus, the processing of dust grain ice-mantles by ionizing radiation is a very real phenomenon and the fact that such interactions can lead to the formation of astrochemically interesting, even prebiotic, species makes radiation chemistry very promising from a modeling standpoint \citep{hudson_amino_2008,holtom_combined_2005,kaiser_theoretical_1997,abplanalp_study_2016}. Until recently, though, the variety and complexity of the underlying microscopic processes have hindered attempts to add radiation-chemical reactions to astrochemical codes \citep{abplanalp_study_2016,shingledecker_cosmic-ray-driven_2018}. 

These processes result from collisions between an incoming particle (the primary ion), and species in the target material. Following the formalism of \citet{bohr_theory_1913}, it is customary to divide such collisions into two categories, namely, those in which energy is transferred to target nuclei (elastic) or electrons (inelastic). One can therefore express the energy loss per unit path length of the primary ion - called the stopping power - as follows:

\begin{equation}
  \frac{\mathrm{d}E}{\mathrm{d}t} \approx n (S_\mathrm{n} + S_\mathrm{e})
\end{equation}

\noindent
where $n$ is the density of the target  (cm$^{-3}$) and $S_\mathrm{n}$ and $S_\mathrm{e}$ are, respectively, the so-called nuclear/elastic and electronic/inelastic stopping cross sections  (eV-cm$^{2}$).  Inelastic collisions, in turn, can further be divided into those in which target species are ionized or excited. Such ionizations result in the formation of so-called ``secondary electrons'' which typically have energies under 50 eV and can themselves ionize or excite target species, thereby further propagating the physicochemical effects initiated by the primary ion \citep{spinks_introduction_1990,johnson_energetic_1990}.

Recently in \citet{shingledecker_general_2018} (hereafter, SH), we introduced a method for treating radiation-chemical processes that is simple enough to include in existing astrochemical models, while simultaneously preserving salient features inferred from experimental data. This method was later added to the astrochemical-type Nautilus v1.1 code \citep{ruaud_gas_2016}, and used by us to determine what, if any, effects these new processes would have on cold core models \citep{shingledecker_cosmic-ray-driven_2018}. Preliminary results from that study indicated that cosmic ray induced reactions could indeed result in enhanced abundances for a number of species, including methyl formate and ethanol, and allowed us to include novel reactions such as  insertions \citep{bergner_methanol_2017,bergantini_mechanistical_2018}. However, given the uncertainties sometimes involved in comparing the results of astrochemical models with observations of interstellar environments - particularly in cases where grain-surface chemistry is involved - questions lingered regarding how well astrochemical models were actually simulating irradiation chemistry.

To that end, we have carried out simulations of two different ice irradiation experiments using an existing astrochemical-type model \citep{vasyunin_formation_2017}, modified to include the SH method for treating radiation chemistry. Specifically, we have modeled the bombardment of a 5 K pure \ce{O2} ice by 100 keV protons as described in \citet{baragiola_solid-state_1999} as well as that of a pure water ice at both 16 K and 77 K by 200 keV protons, as reported in \citet{gomis_hydrogen_2004-1}. Doing so has not only afforded us the opportunity to compare how well our simulations replicate the experimental data, but also to directly gauge how well our models perform, more generally, with handling reactions in the bulk ice over a range of temperatures. By thus testing our approaches to ice chemistry in well constrained experimental systems, confidence can be increased in the results of models of much less well constrained interstellar environments. 

The rest of the paper is as follows: in \S \ref{sec:model} we discuss details of both the model and chemical network used in this work. In \S \ref{sec:results} our major results are described and their significance is discussed. Finally, our conclusions and suggestions for future work are presented in \S \ref{sec:conclusions}.

\section{Model and Network} \label{sec:model}

The basis of this work is the \texttt{MONACO} code, described in \citet{vasyunin_formation_2017}, which uses a rate-equation approach for reactions in the gas phase, and the so-called ``modified rate-equation'' method  for reactions on and in dust grain ice mantles \citep{garrod_new_2008,garrod_new_2009}. As in \citet{vasyunin_unified_2013} we assume the selvedge,  a term for the altered layers near the top of the mantle,  is comprised of the top four monolayers of the ice \citep{vasyunin_unified_2013}.  

We have modified the code to account for irradiation-chemical reactions on and in ice using the SH method, the foundation of which is the assumption that one of the following outcomes is possible upon collision between an energetic particle, the primary ion or a secondary electron, and some target species, $A$:

\begin{equation}
A  \leadsto A^+ + e^- 
\tag{P1}
\label{p1}
\end{equation}

\begin{equation}
A  \leadsto A^+ + e^- \rightarrow A^* \rightarrow B^* + C^* 
\tag{P2}
\label{p2}
\end{equation}

\begin{equation}
A  \leadsto A^* \rightarrow B + C 
\tag{P3}
\label{p3}
\end{equation}

\begin{equation}
A  \leadsto A^* 
\tag{P4}
\label{p4}
\end{equation}

\noindent
where the curly arrow  indicates a collision. In processes (P1) and (P2), $A$ is ionized, though with fast charge recombination leading to the formation of suprathermal dissociation products, designated with an asterisk, assumed to occur in \eqref{p2}. Processes (P3) and (P4) represent collisions in which a suprathermal product is electronically excited directly into some higher state which, in the case of (P3), is assumed to lead to the formation of thermal products. The efficiencies of (P1) - (P4) are characterized by radiochemical yields, called $G$-values, which give the number of molecules created (or destroyed) per 100 eV transferred to the system from bombarding particles \citep{dewhurst_general_1952,spinks_introduction_1990}. A full list of the radiolysis processes used here is given in Table \ref{tab:gvalues} of Appendix \ref{sec:radiolysis}. The yields given there were arrived at through a two-step process where first, initial values were estimated usnig the SH method and subsequently adjusted to improve agreement with the experimental data.
It should be noted that, following \citet{shingledecker_cosmic-ray-driven_2018}, we assume the rate of process \eqref{p1} is zero, i.e. that all charged species produced in the bulk via ionization are quickly neutralized. 

First-order rate coefficients for each radiolysis process are calculated using

\begin{equation}
	k_\mathrm{rad} = \left(\frac{G}{100\;\mathrm{eV}}\right) S_\mathrm{e}  \phi,
   \label{krad}
\end{equation} 

\noindent
with $\phi$ being the radiation flux and $S_\mathrm{e}$ the previously described inelastic/electronic stopping cross section. In this work, we assume the flux to be $1.0\times10^{11}$ \ce{H+} $\mathrm{cm}^{-2}$ $\mathrm{s^{-1}}$, and use the stopping cross sections given in Table \ref{tab:experiment}.  

We assume that, in the bulk, suprathermal species produced via radiolysis rapidly undergo either (a) reaction with a neighbor, or (b) are quenched by the surrounding material. Rate coefficients for the reaction between $A$ and $B$ - where either one could be suprathermal - are calculated using the expression

\begin{equation}
  k_\mathrm{fast} = 
  f_\mathrm{br}
  \left[
  \frac{
  \nu_0^A + 
  \nu_0^B
  }
  {
  {N_\mathrm{bulk}}
  }
  \right]
  \mathrm{exp}
  \left(-
  \frac 
  {E_\mathrm{act}^{AB}}
  {T_\mathrm{ice}}
  \right),
  \label{ksup}
\end{equation}

\noindent
where  $f_\mathrm{br}$ is the branching fraction, $N_\mathrm{bulk}$ is the total number of bulk species in the simulated ice,  
and $E_\mathrm{act}^{AB}$ is the activation energy for reaction. Absent other data, we assume here that suprathermal species react barrierlessly. The characteristic vibrational frequency, $\nu^A_0$, is given by

\begin{equation}
  \nu_{0}^{A} = \sqrt{
  \frac
  {2n_\mathrm{s}E^A_\mathrm{D}}
  {\pi^2m_A}
  },
  \label{trialnu}
\end{equation}

\noindent
with  $E^A_\mathrm{D}$ and $m_A$ being the desorption energy  and mass of $A$, respectively, and $n_\mathrm{s}$ the phsysisorption site density - here $6.55\times10^{14}$ cm$^{-2}$ \citep{vasyunin_formation_2017}. This characteristic vibrational frequency is also used as the first-order rate coefficient for the quenching of the suprathermal species by the ice. It should be noted that, particularly in water ice, these electronic excitations, also called excitons, may diffuse from the interior of the ice to the selvedge, where they can drive the desorption of species into the gas \citep{thrower_highly_2011,marchione_efficient_2016}. Thus, our assumption that these suprathermal species are rapidly quenched should be seen as a first-order approximation of their real behavior.

As an example of how the above equations are utilized in our code, consider the time-dependent abundance of $B^*$. Assuming $B^*$ is produced via the radiolysis of $A$ as in \eqref{p2}, and destroyed both via quenching and reaction with some other bulk species, $X$, the rate of change in the number of $B^*$ as a function of time can be described by

\begin{equation}
    \frac{d\,N_{B^*}}{dt} = k_\mathrm{rad}^\mathrm{\ref{p2}}N_A - \nu_0^B N_{B^*} - k_\mathrm{fast}N_X N_{B^*},
\end{equation}

\noindent
where the reaction between $B^*$ and $X$ is, as with all suprathermal species here, assumed to occur barrierlessly.

\begin{deluxetable*}{lcc}
\tablecaption{Physical parameters used in simulations of experiments. \label{tab:experiment}}
\tablehead{
\colhead{Parameter} & \colhead{Value} & \colhead{Source}  
}
\startdata
\multicolumn{3}{c}{\ce{O2}} \\ \\
Temperature     & 5 K   & \citet{baragiola_solid-state_1999} \\
Density         & $5.7\times10^{22}$ cm$^{-3}$ & \citet{horl_structure_1982} \\
\ce{H+} Energy & 100 keV &  \citet{baragiola_solid-state_1999} \\
$S_\mathrm{e}$  &  $4.6\times10^{-14}$ eV cm$^2$ & \citet{baragiola_solid-state_1999} \\ \hline
\multicolumn{3}{c}{\ce{H2O}} \\ \\
Temperature  & 16 K, 77 K & \citet{gomis_hydrogen_2004} \\
Density      & $3.1\times10^{22}$ cm$^{-3}$ & \citet{martonak_evolution_2005} \\
\ce{H+} Energy & 200 keV & \citet{gomis_hydrogen_2004} \\
$S_\mathrm{e}$  &  $3.3\times10^{-14}$ eV cm$^2$ & \citet{uehara_calculations_2000} \\ \hline
\enddata
\end{deluxetable*}

In order to gauge how well astrochemical models are able to reproduce bulk chemistry, we have run the following three sets of simulations for each experiment considered here using the parameters listed in Table \ref{tab:experiment}:

\begin{enumerate}[label=(\alph*)]
    \item a set of models in which thermal radicals and atomic oxygen were assumed to react non-diffusively with neighboring species in the ice with rate coefficients calculated using Eq. \eqref{ksup},
    \item a set of models in which all bulk rate coefficients were calculated using the standard diffusive formula \citep{hasegawa_models_1992} along with diffusion barriers obtained from the desorption energies listed in Table \ref{tab:desorption} and no tunneling under diffusion barriers for any species, and 
    \item a set of models similar to (b) but with tunneling through diffusion barriers allowed for H, \ce{H2}, and O.
\end{enumerate}

\noindent

For the third set of simulations, we treat tunneling under diffusion barriers by H and \ce{H2} using the standard formalism in \citet{hasegawa_models_1992} with a barrier width of 1.0 \AA. For atomic oxygen, following results from \citet{minissale_quantum_2013}, we use a desorption energy of 1040 K and a barrier width of 0.7 \AA. In all simulations, H and \ce{H2} were assumed to tunnel through reaction barriers \citep{tielens_model_1982}. The competition mechanism for systems with chemical activation energies \citep{chang_gas-grain_2007,herbst_chemistry_2008} is used here in all models only for reactions on the surface.

\begin{deluxetable*}{lcc}
\tablecaption{Desorption energies for relevant species. \label{tab:desorption}}
\tablehead{
\colhead{Species} & \colhead{$E_\mathrm{D}$ (K)} & \colhead{Source} 
}
\startdata
\ce{O}  & 1\,660\tablenotemark{a} & \citet{he_new_2015} \\
        &  1\,040\tablenotemark{b} & \citet{minissale_quantum_2013} \\
\ce{O2}  & 930 & \citet{jing_formation_2012} \\
\ce{O3}  & 1\,833  & \citet{jing_formation_2012} \\
\ce{H} & 450  & \citet{garrod_formation_2006} \\\citep{thrower_highly_2011,marchione_efficient_2016}
\ce{H2} & 430  & \citet{garrod_formation_2006} \\
\ce{OH} & 2\,850  & \citet{garrod_formation_2006} \\
\ce{H2O} & 5\,700  & \citet{garrod_formation_2006} \\
\ce{HO2} & 4\,510  & $E_\mathrm{D}^\mathrm{O} + E_\mathrm{D}^\mathrm{OH}$ \\
\ce{H2O2} & 5\,700  & $E_\mathrm{D}^\mathrm{OH} + E_\mathrm{D}^\mathrm{OH}$ \\
\enddata
\tablenotetext{a}{Used in models (a) and (b)}
\tablenotetext{b}{Used in model (c)}
\tablecomments{Diffusion barriers in the selvedge were assumed to be $0.5\times E_\mathrm{D}$, while those in the bulk to be $0.7\times E_\mathrm{D}$.}
\end{deluxetable*}

For the chemical network, we used the reactions listed in Table \ref{tab:h2onetwork} of Appendix \ref{sec:network}, as well as the neutral-neutral oxygen reactions listed in Table 7 of \citet{shingledecker_new_2017}. In cases where we have been unable to find values for activation energies, we have assumed a barrier of $10^4$ K, except for radical-radical reactions, which were assumed to be barrierless. For every thermal reaction of the type given in Table \ref{tab:h2onetwork}, we included variants involving one suprathermal reactant. To illustrate this point, for every reaction of the form $\mathrm{A + B \rightarrow} \; products$, we include the following suprathermal variants

\begin{equation}
    \mathrm{A^* + B \rightarrow} \; products
    \tag{R1}
    \label{r1}
\end{equation}

\begin{equation}
    \mathrm{A + B^* \rightarrow} \; products.
    \tag{R2}
    \label{r2}
\end{equation}

\noindent
A number of reactions listed in Table \ref{tab:h2onetwork} have barriers that are quite high for astrochemical networks,
but are more common in the combustion-chemical literature from which most are taken. Even though the rates of such reactions involving thermal species will indeed be negligible in our models, the variants involving suprathermal species should be efficient even at the temperatures considered here. For example, even a $40\,000$ K barrier corresponds to only $\sim3.4$ eV, a realistic energy for an electronically excited suprathermal species. 

\section{Results and Discussion} \label{sec:results}

In order to examine how well astrochemical models using the SH method are capable of reproducing experimental data, as well as to determine the accuracy of bulk chemistry in such codes, we have simulated ice-irradiation systems described in two previous radiation-chemical studies discussed above - the results of which are described below.

\subsection{H$^+$ Bombardment of Pure O$_2$ Ice}

\begin{figure*}[ht!]
\gridline{
          \fig{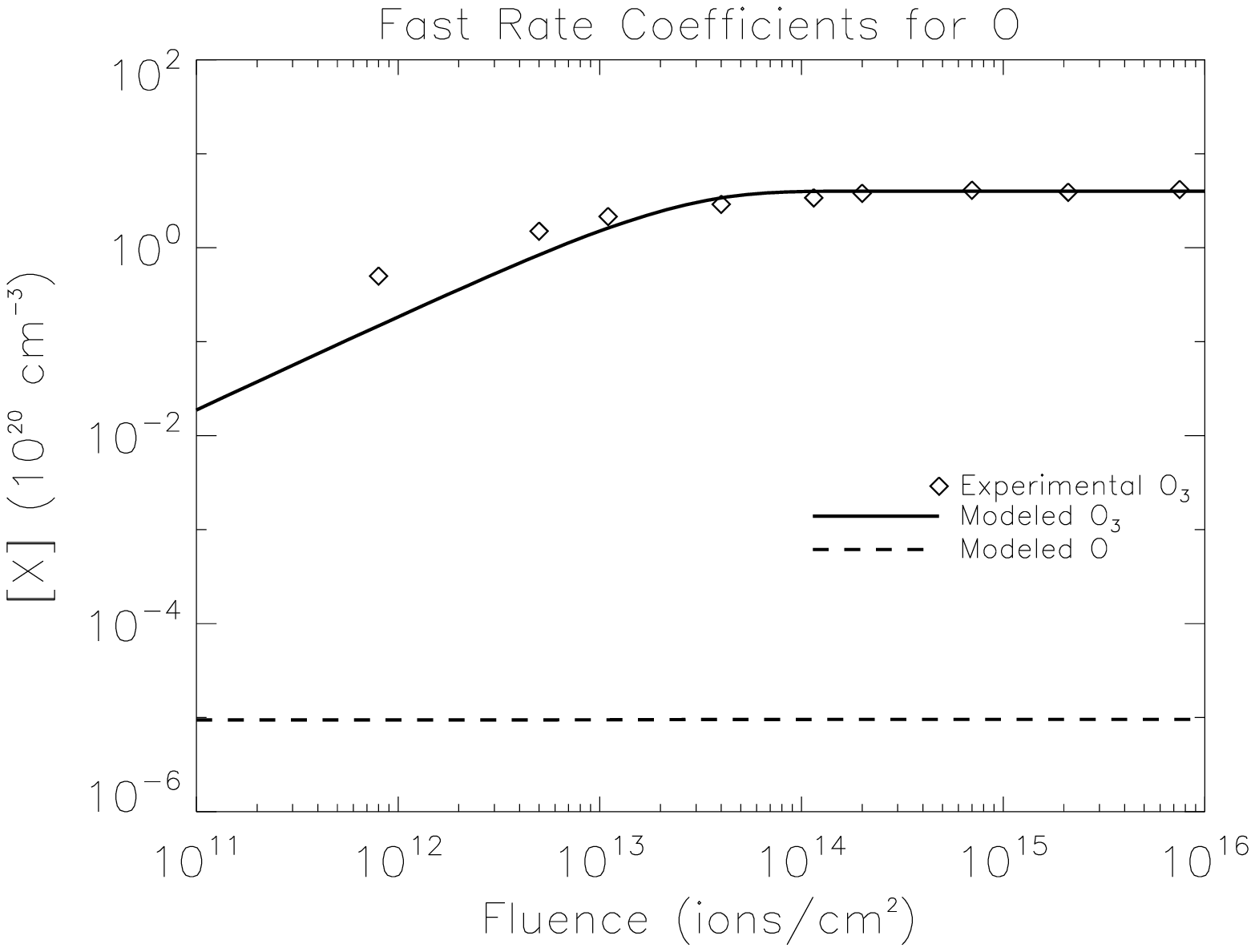}{0.45\textwidth}{(a)}
         }
\gridline{
          \fig{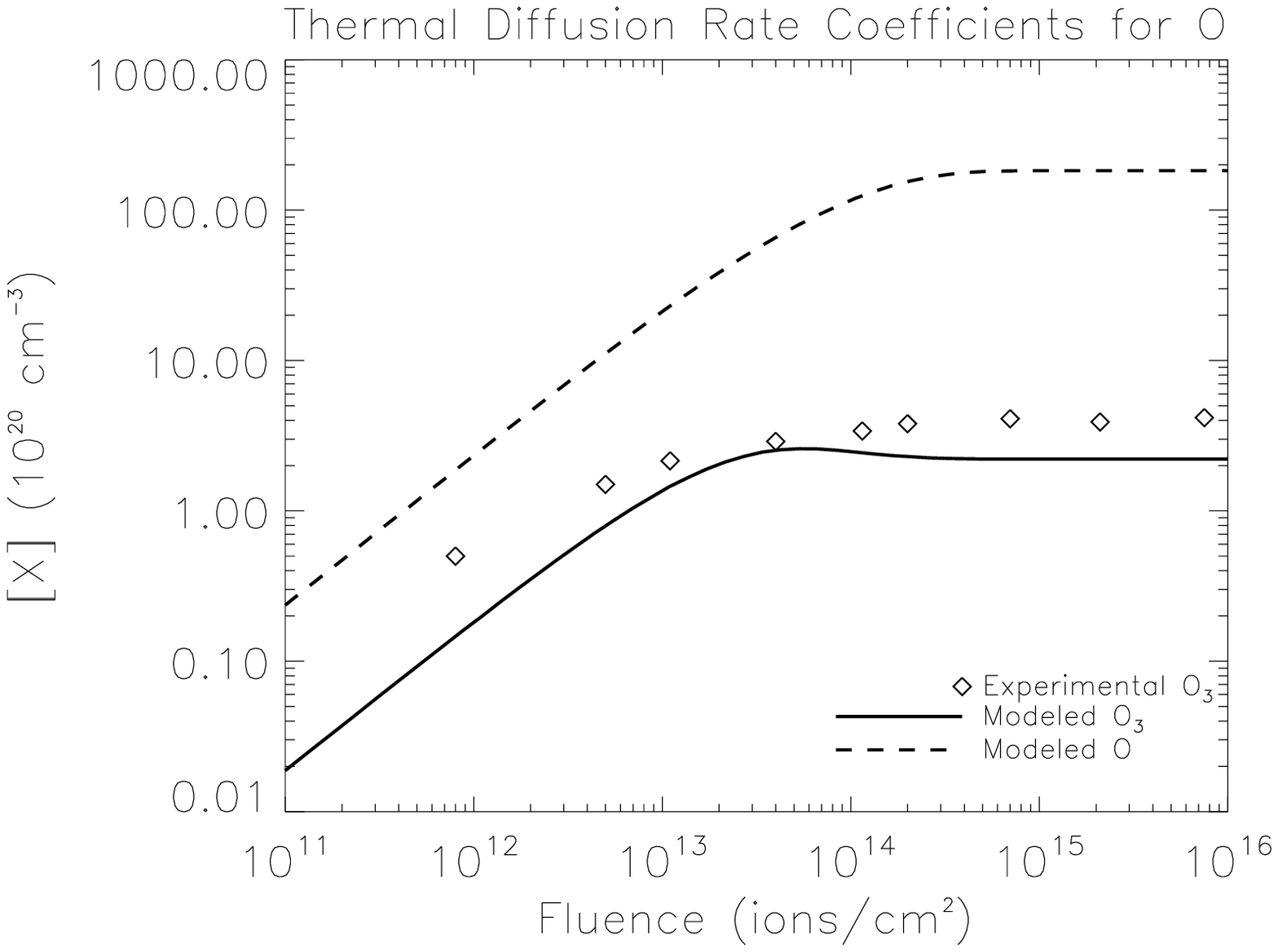}{0.45\textwidth}{(b)}
          \fig{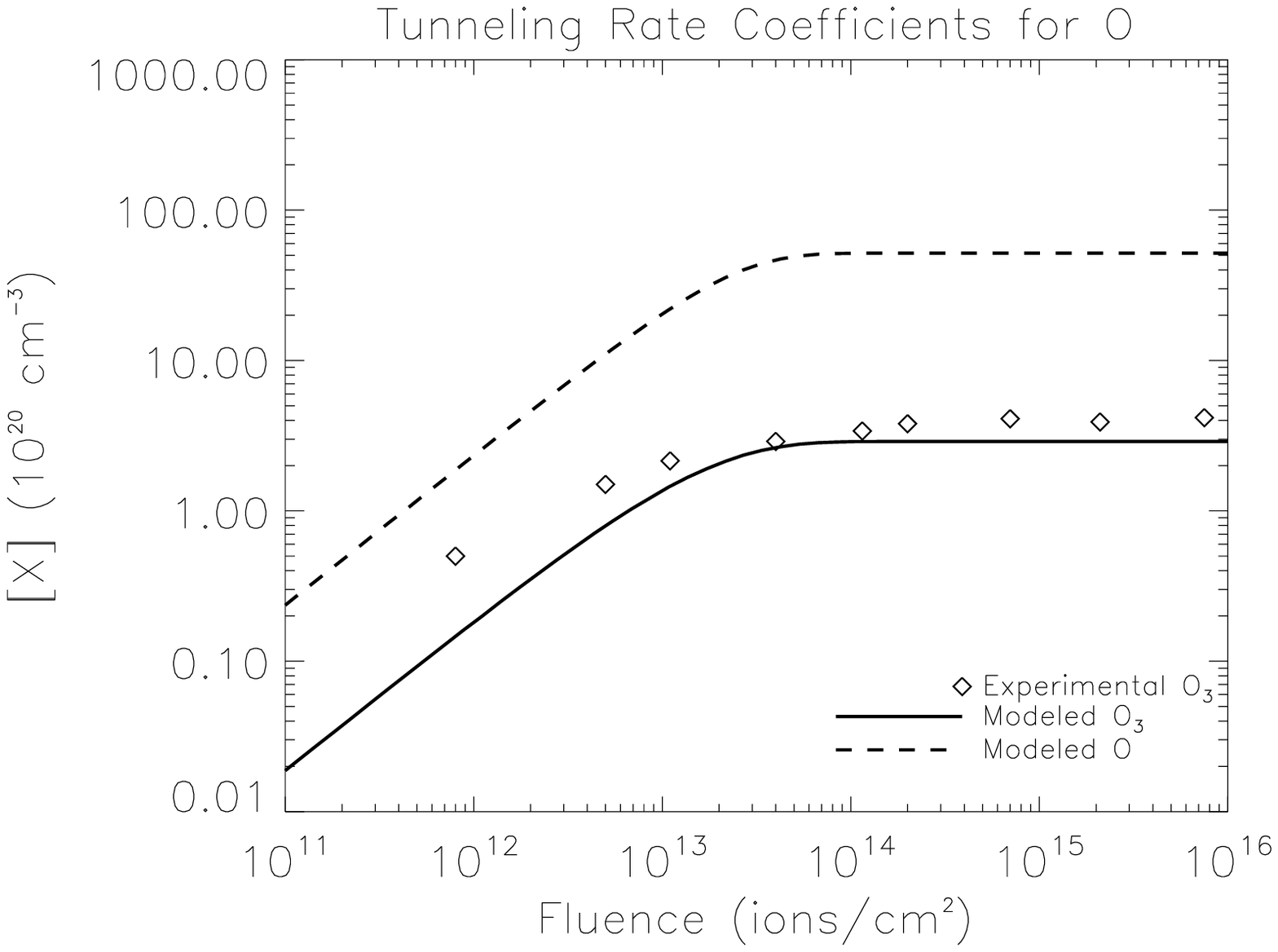}{0.45\textwidth}{(c)}
         }
\caption{Results from simulations of an irradiated pure O$_2$ ice using (a) fast bulk reaction of O, (b) standard bulk rate coefficients based on thermal hopping, and (c) O atom tunneling. 
\label{fig:o2ice}}
\end{figure*}

In this system, the only radical used is atomic oxygen, which can be thermal or suprathermal. We first simulated the irradiation of a pure \ce{O2} ice at 5 K by 100 keV \ce{H+}, following the study by \citet{baragiola_solid-state_1999}, which has previously been successfully modeled by us using the much more detailed Monte Carlo code, \texttt{CIRIS} \citep{shingledecker_new_2017}. An irradiated \ce{O2} ice is, in many respects, an ideal system for testing simulations of radiation chemistry, given the limited number of possible neutral species, i.e. O, \ce{O2}, and \ce{O3}. However, in the Monte Carlo code even this seemingly simple system required a network of $\sim50$ reactions or processes and involved a level of detail beyond the practical capabilities of rate equations-based astrochemical codes, such as explicitly calculating the tracks of incident ions and secondary electrons or following each ice species on a hop-by-hop basis.  

Despite these limitations though,  as shown in Fig. \ref{fig:o2ice}a, even our simplified approach is capable of reasonably reproducing the concentration of  \ce{O3} as a function of fluence. Here, fluence - the product of the \ce{H+} flux and irradiation time - represents the number of ions that have bombarded the ice per cm$^2$. As can be seen in Fig. \ref{fig:o2ice}, the best results were obtained in model (a), where atomic oxygen is assumed to react non-diffusively in the bulk, yielding excellent agreement between calculated and experimental abundances in the steady stage regime. An analogous assumption was made in our previous Monte Carlo simulations using \texttt{CIRIS}, with similarly good results  \citep{shingledecker_new_2017}. The slight under-prediction of \ce{O3} at fluences under $\sim10^{13}$ \ce{H+} cm$^{-1}$ is likely due to additional processes - such as dissociative electron attachment (DEA) - which are also driven by energetic ion bombardment but not currently considered in this simplified treatment of radiolysis \citep{arumainayagam_low-energy_2010}. Also noticeable in Fig. \ref{fig:o2ice}a is the fact that atomic oxygen abundances remained low throughout the simulation, despite its continuous production from the dissociation of \ce{O2} and, to a lesser degree, \ce{O3}. 

Comparison with Fig. \ref{fig:o2ice}b reveals that model (b), using standard Langmuir-Hinshelwood rate coefficients, is the least accurate. There, though the modeled \ce{O3} abundance is only a factor of a few lower than the experimental values, an unrealistically large abundance of atomic oxygen is predicted, contrary to what has been suggested by previous studies  \citep{gerakines_ultraviolet_1996,baragiola_solid-state_1999,bennett_laboratory_2005}. This buildup is caused by the very slow rates of thermal diffusion reactions at 5 K, given the high atomic oxygen desorption energy of 1660 K \citep{he_new_2015}. 

In order to determine the possible effects of quantum tunneling by atomic oxygen through diffusion barriers, as suggested by \citet{minissale_quantum_2013}, we ran further simulations using their best-fit results of a barrier width of $0.7$ \AA $\;$ and a desorption energy of 1040 K (assuming $E_\mathrm{b} = 0.5 E_\mathrm{D}$). Tunneling for H and \ce{H2} is treated using the method of \citet{hasegawa_models_1992}, in which a barrier width of 1 \AA$\;$ is assumed.
Comparison of the results of this model, shown in Fig. \ref{fig:o2ice}c, with \ref{fig:o2ice}b reveals that though agreement between the calculated and experimental ozone abundances is improved, the abundance of atomic oxygen remains too high.

The results of these three models demonstrate that best agreement with experimental data - both the measured \ce{O3} abundance as well as the inferred low O abundance - is obtained when atomic oxygen is assumed to react quickly in the bulk. We note, however, that the predicted atomic oxygen abundances in model (a) may in fact be somewhat too low, based on the observation by \citet{bennett_laboratory_2005} of an IR feature associated with an [\ce{O3}...O] complex, though no estimate of the atomic oxygen abundance (or an upper limit) was derived. In such irradiated ices, it is likely that some fraction of the oxygen atoms cannot react quickly, either due to trapping by \ce{O3} or steric effects, in which case their abundance would indeed be greater than that predicted in our models. These effects could be accounted for in the method used here by decreasing the value of the pre-exponential frequency, $\nu_0$, thereby increasing the abundance of O. Here, we use the characteristic vibrational frequency given in Eq. \eqref{trialnu} (which typically has a value on the order of $10^{12}$ s$^{-1}$) as a rough approximation, though in fact, the value of this parameter can be as low as $10^{-3}$ s$^{-1}$ \citep{theule_thermal_2013}, and rate coefficients obtained using \eqref{trialnu} should probably be interpreted as upper limits.

\subsection{H$^+$ Bombardment of Pure H$_2$O Ice}

We next simulated the irradiation of a pure \ce{H2O} ice by 200 keV \ce{H+} at both 16 and 77 K 
using the three different bulk chemistry schemes described in Sec. \ref{sec:model}. Here, in addition to atomic oxygen, we further assume that all radicals in our network -- H, OH, and \ce{HO2} -- react quickly in model (a). This system, though substantially more complex than \ce{O2}, is of greater astronomical interest, given the ubiquity of water ice in planetary bodies \citep{altwegg_67p/churyumov-gerasimenko_2015,carlson_europas_2009}, as well as interstellar dust grain ice mantles \citep{gibb_interstellar_2004}. 

\begin{figure*}[p]
\gridline{
          \fig{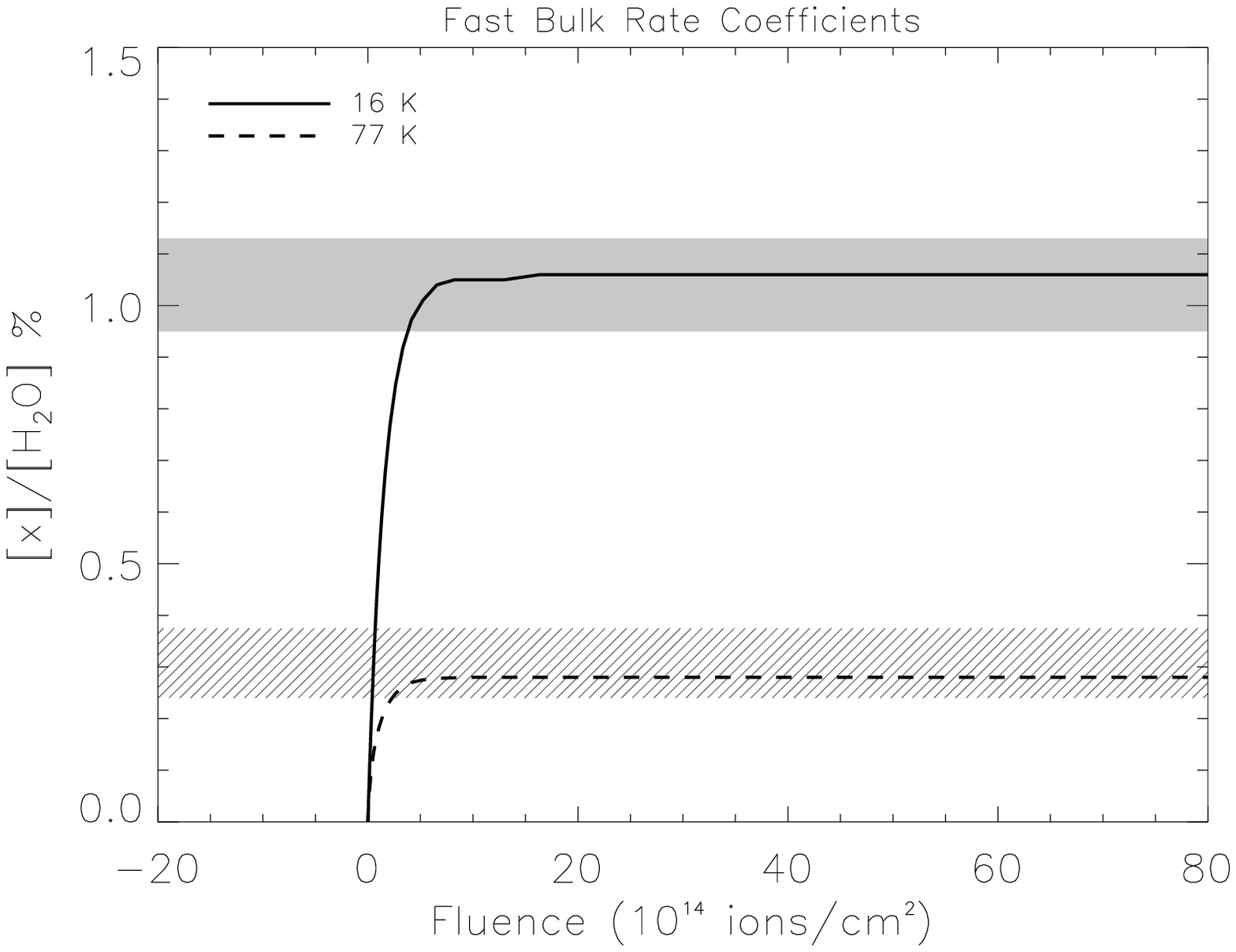}{0.45\textwidth}{(a)}
          }
\gridline{
          \fig{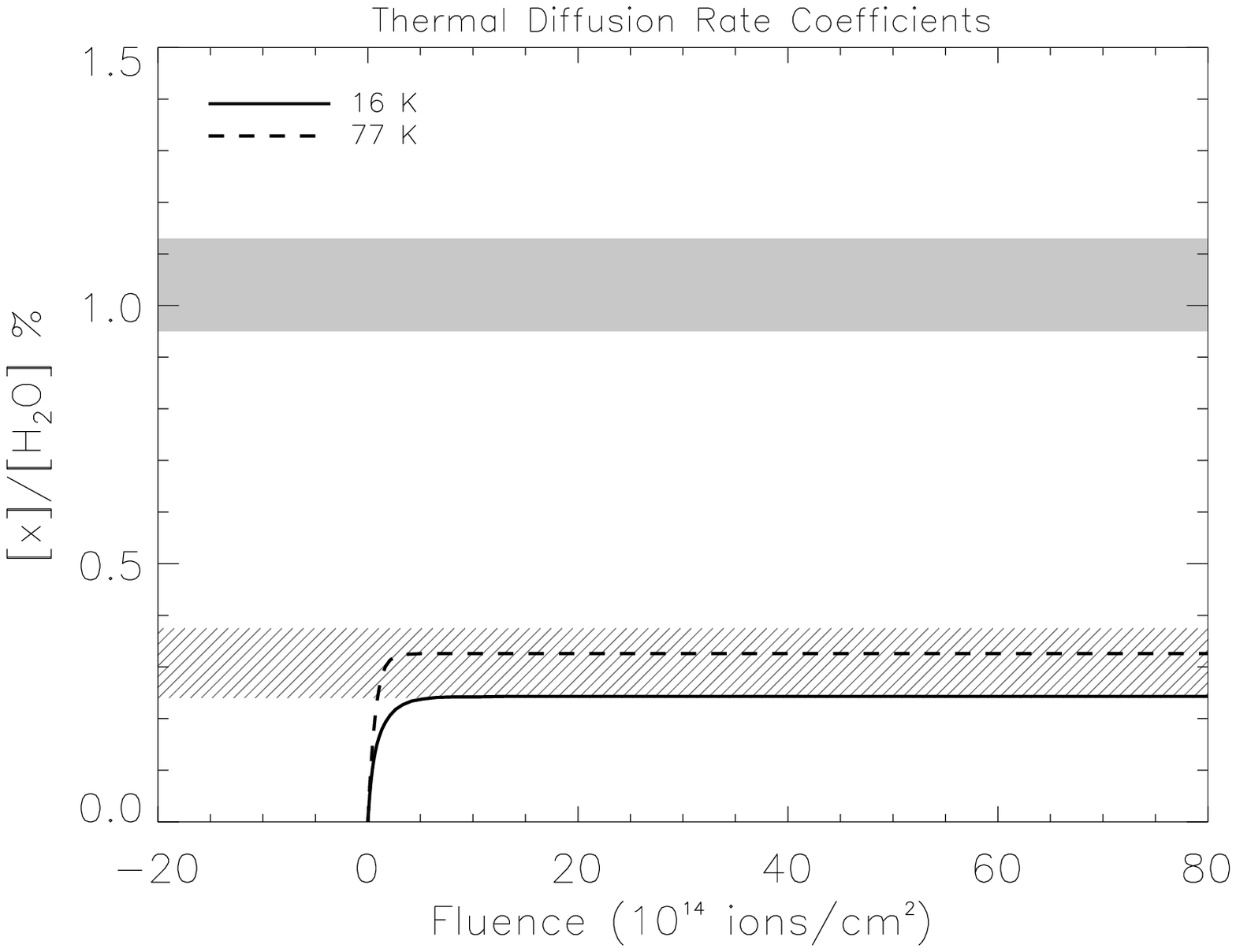}{0.45\textwidth}{(b)}
          \fig{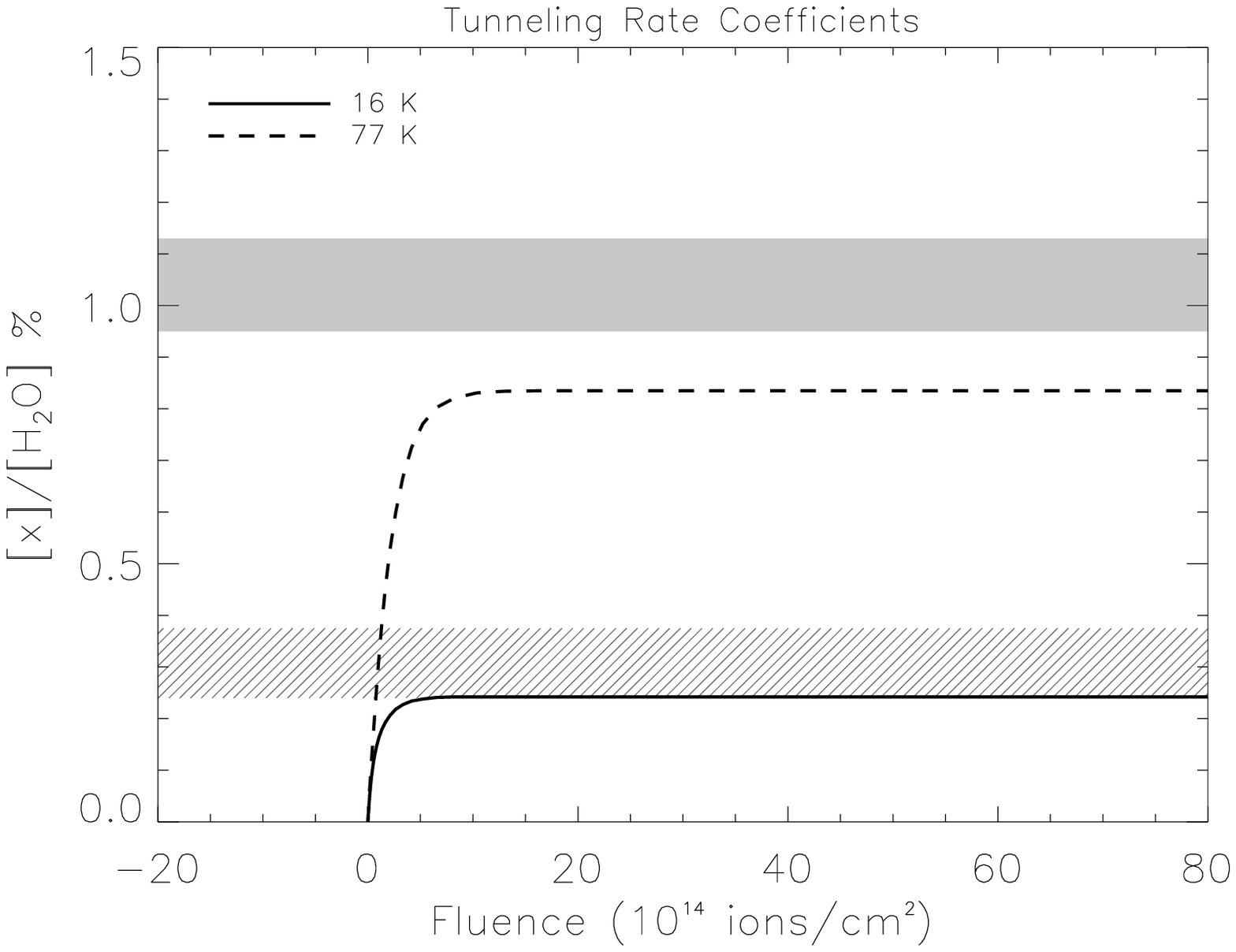}{0.45\textwidth}{(c)}
          }
\caption{
Calculated abundances of \ce{H2O2} versus proton fluence from simulations of a pure \ce{H2O} ice bombarded by 200 keV \ce{H+} assuming (a) fast bulk reactions of radicals and atomic oxygen, (b) only thermal bulk diffusion, and (c) diffusion barrier tunneling for H, \ce{H2}, and O. Approximate steady-state hydrogen peroxide abundances from \citet{gomis_hydrogen_2004-1} at both 16 K and 77 K are represented by the solid, and line-filled boxes, respectively. 
\label{fig:relative}}
\end{figure*}

Shown in Fig. \ref{fig:relative} are both our calculated \ce{H2O2} abundances vs proton fluence as well as the approximate steady-state bf experimental values from \citet{gomis_hydrogen_2004-1} relative to \ce{H2O} of $\sim 1.0$\% at 16 K and $0.25$\% at 77 K. As with the pure \ce{O2} ice, model (a) once again provides the best agreement with experimental data and, notably, is the only one of the three to predict a drop in hydrogen peroxide abundance at higher temperatures. \citet{moore_ir_2000}, who first detected this trend, speculated that it was due to the reaction

\begin{equation}
  \ce{OH + H2O2 -> H2O + HO2}
  \tag{R3}
  \label{r3}
\end{equation}

\noindent

\begin{figure*}[p]
 \gridline{
          \fig{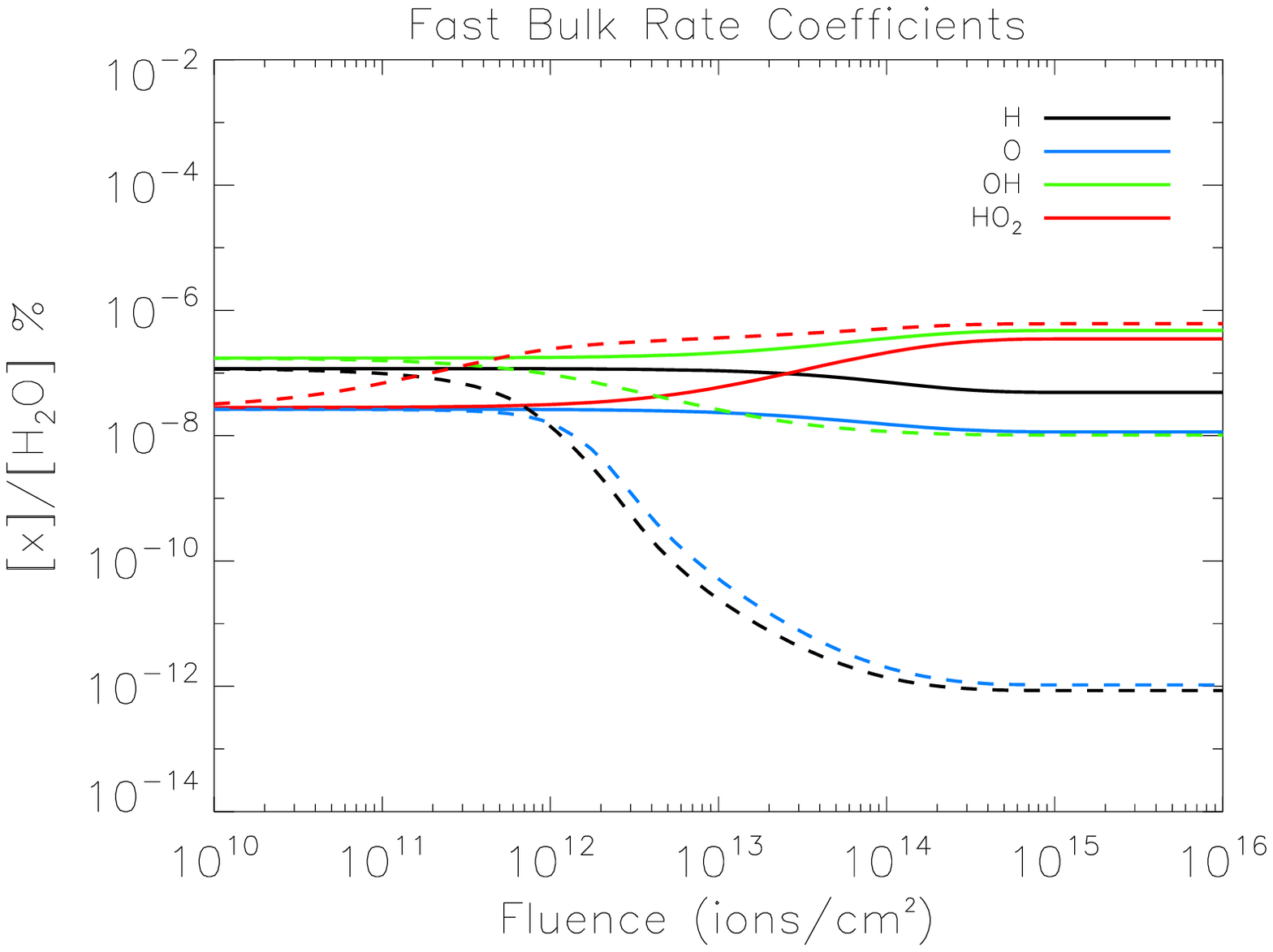}{0.45\textwidth}{(a)}
          }         
 \gridline{
          \fig{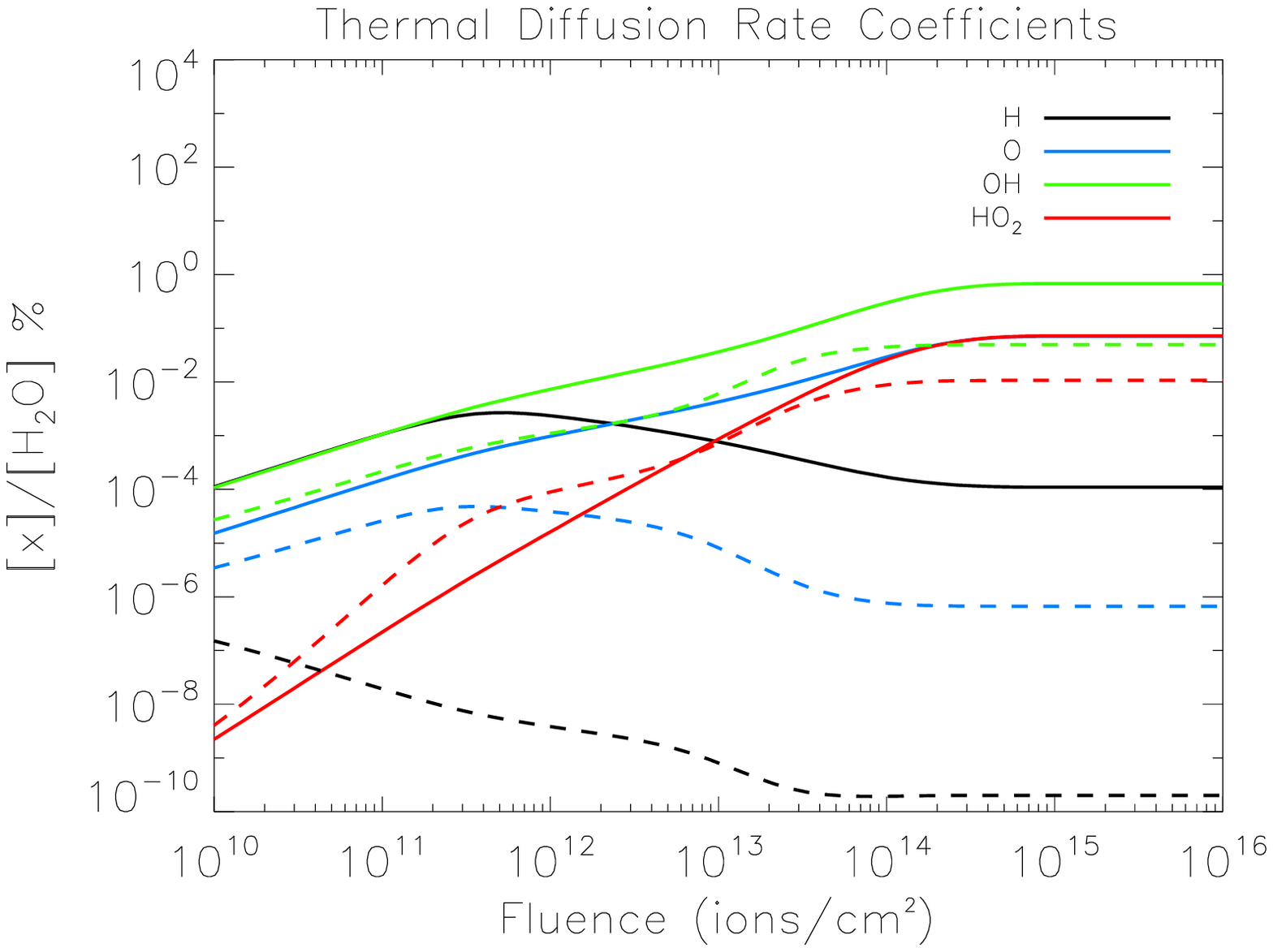}{0.45\textwidth}{(b)}
          \fig{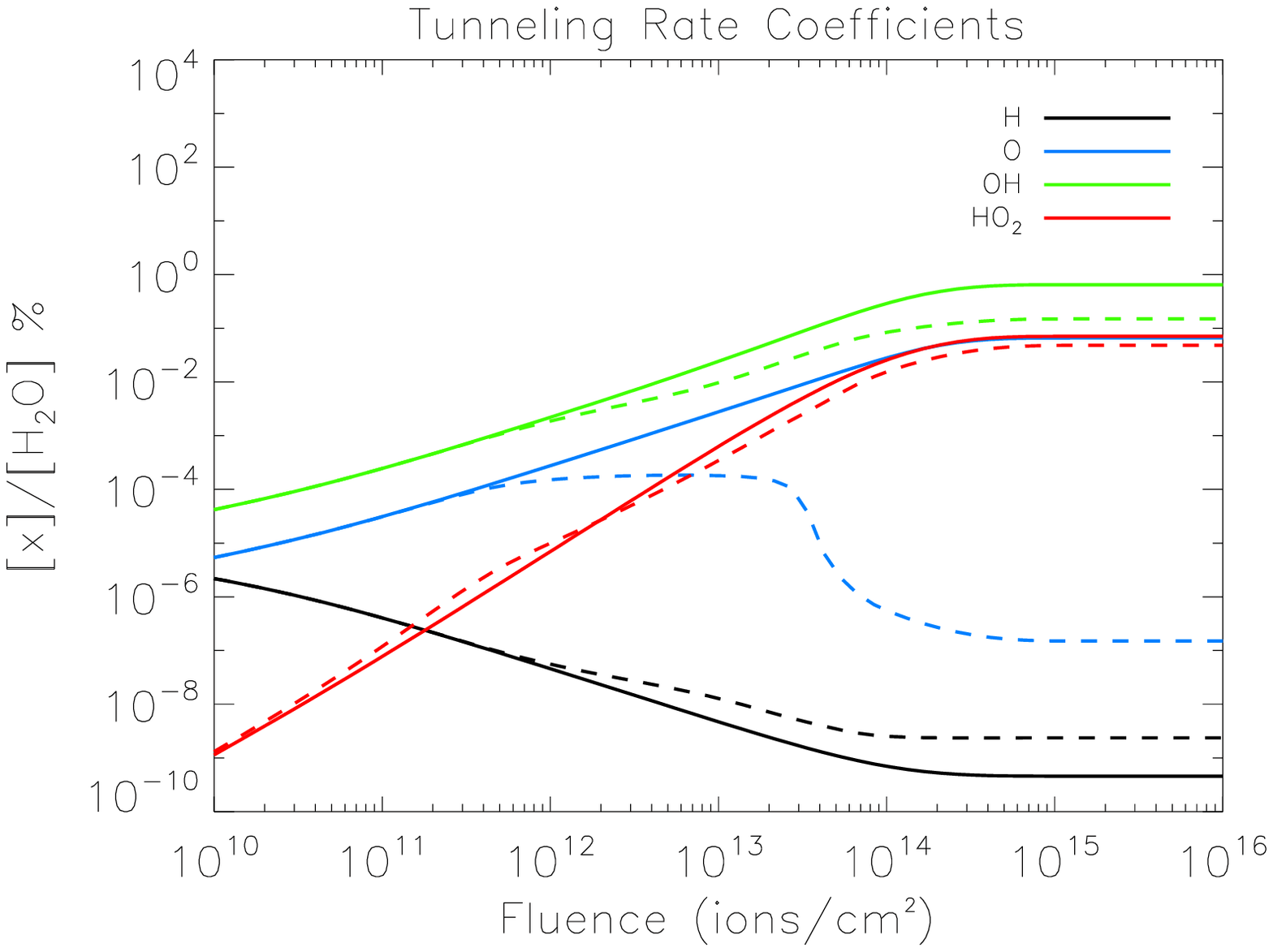}{0.45\textwidth}{(c)}
          }         
\caption{
Calculated abundances of \ce{H}, \ce{O}, \ce{OH}, and \ce{HO2} versus proton fluence at 16 K (solid lines) and 77 K (dashed lines) from simulations of a pure \ce{H2O} ice bombarded by 200 keV \ce{H+} assuming (a) fast bulk reactions of radicals and atomic oxygen, (b) only thermal bulk diffusion, and (c) diffusion barrier tunneling for H, \ce{H2}, and O. 
\label{fig:radicals}}
\end{figure*}

We find that, in agreement with \citet{moore_ir_2000}, reaction \eqref{r3} is indeed behind the drop in \ce{H2O2} abundance at $\sim 80$ K, since, even though it is assumed that radicals such as OH react quickly, \eqref{r3} only becomes competitive at 77 K due to the 755.3 K barrier \citep{ginovska_reaction_2007}. In models (b) and (c) the effect of reaction \eqref{r3} is  reduced by the slow diffusion rates of OH and \ce{H2O2}. As can be seen in Fig. \ref{fig:radicals}, without the assumption that radicals react quickly in the bulk, OH, in particular, becomes quite abundant in the ice, even at 77 K.

The increase in hydrogen peroxide abundance at 77 K in models (b) and (c)  is due to the following reactions:

\begin{equation}
  \ce{OH + OH -> H2O2}
  \tag{R4}
  \label{r4}
\end{equation}
\begin{equation}
  \ce{H + HO2 -> H2O2},
  \tag{R5}
  \label{r5}
\end{equation}

\noindent
where here, the higher rate of \ce{H2O2} formation is due to the increased mobility of the reactants. Enabling quantum tunneling through diffusion barriers, as in model (c), further speeds up the rate of reaction \eqref{r5}, as well as the formation of the precursor species, \ce{HO2}, via 

\begin{equation}
  \ce{H + O2 -> HO2},
  \tag{R6}
  \label{r6}
\end{equation}

\noindent
thereby contributing to the even higher \ce{H2O2} abundance at 77 K in model (c) compared with model (b).  

\begin{figure*}[ht!]
          \fig{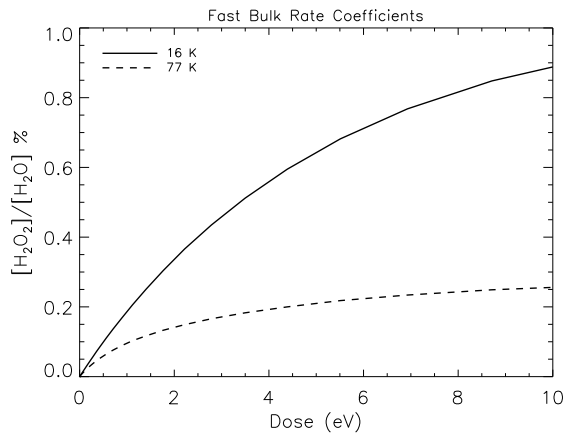}{0.45\textwidth}{}
\caption{
Percentage \ce{H2O2} vs. dose for model (a) in which radicals are assumed to react quickly. Following \citet{moore_ir_2000}, based on the slopes of linear fits to these data, we estimate equivalent measured $G(\ce{H2O2})$-values of 0.1 and 0.03 molecules/100 eV for the 16 K and 77 K simulations, respectively. 
\label{fig:doseplot}}
\end{figure*}

We can gain further insights into how closely our simulations are replicating the experiment by examining $G$(\ce{H2O2}), the radiolytic hydrogen peroxide formation yield.
Unfortunately, we cannot simply compare the $G$-values given in Table \ref{tab:gvalues} with experimentally determined ones directly,
since our values are more representative of the immediate creation (or destruction) of target species, i.e. the efficiencies of each of the microscopic radiolytic processes given in \eqref{p1}-\eqref{p4}, than the single effective experimental value, which is sensitive to the temperature-dependent chemistry of the system \citep{spinks_introduction_1990}. However, following the method used in \citet{moore_ir_2000}, we can estimate what the experimental $G$-value might be from the slope of a linear fit to the abundance curves in Fig. \ref{fig:doseplot} over the pre-steady-state regime - corresponding to doses of ca. 0-10 eV, where dose is the product of the fluence and $S_\mathrm{e}$. From this, we calculate the yield of H$_2$O$_2$ at 16 K to be $0.1$ molecules/100 eV - exactly what was mentioned by \citet{moore_ir_2000} as the yield in pure \ce{H2O}.

\begin{figure*}[p]
\gridline{
          \fig{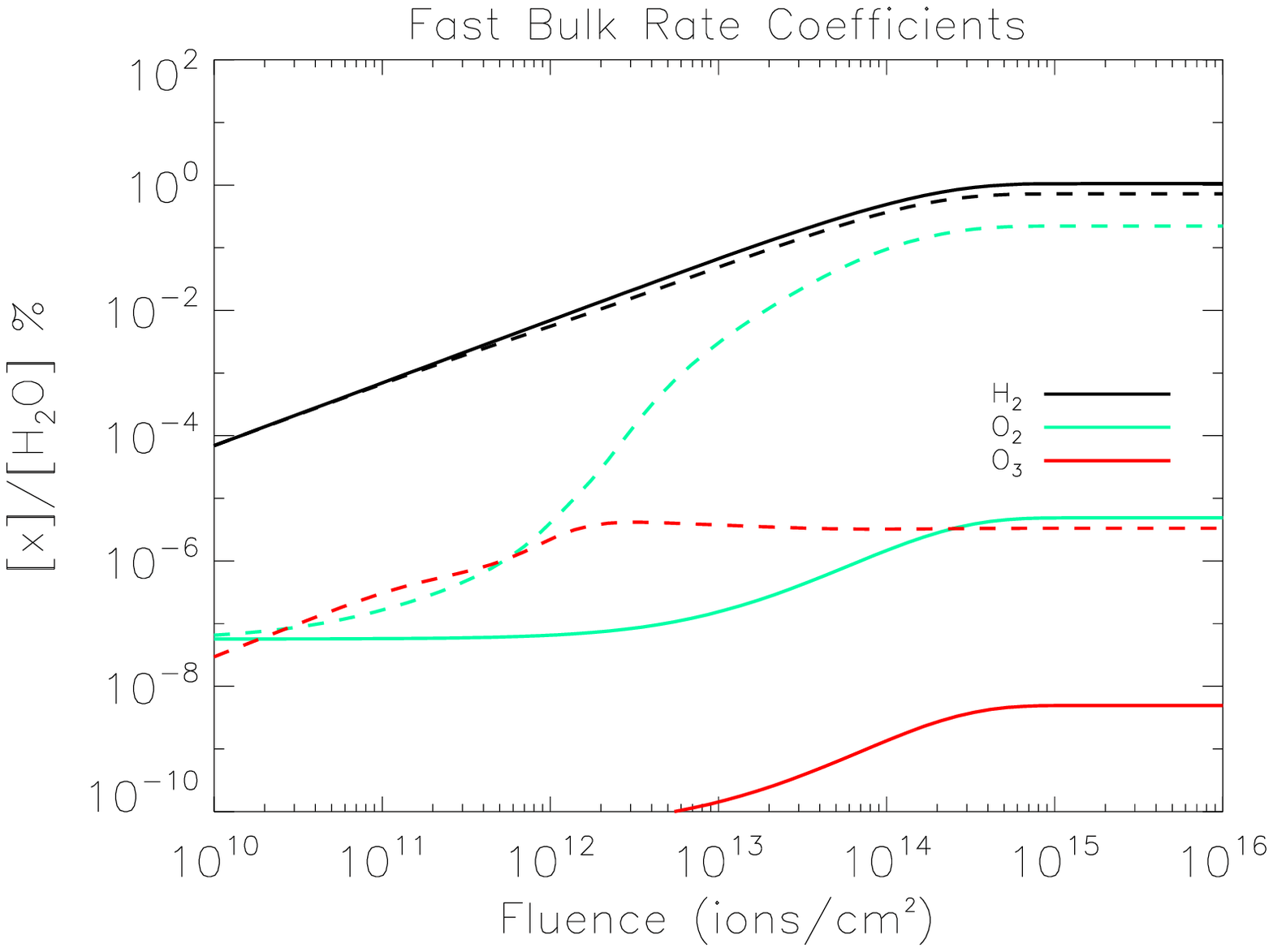}{0.45\textwidth}{(a)}
          }
\gridline{
          \fig{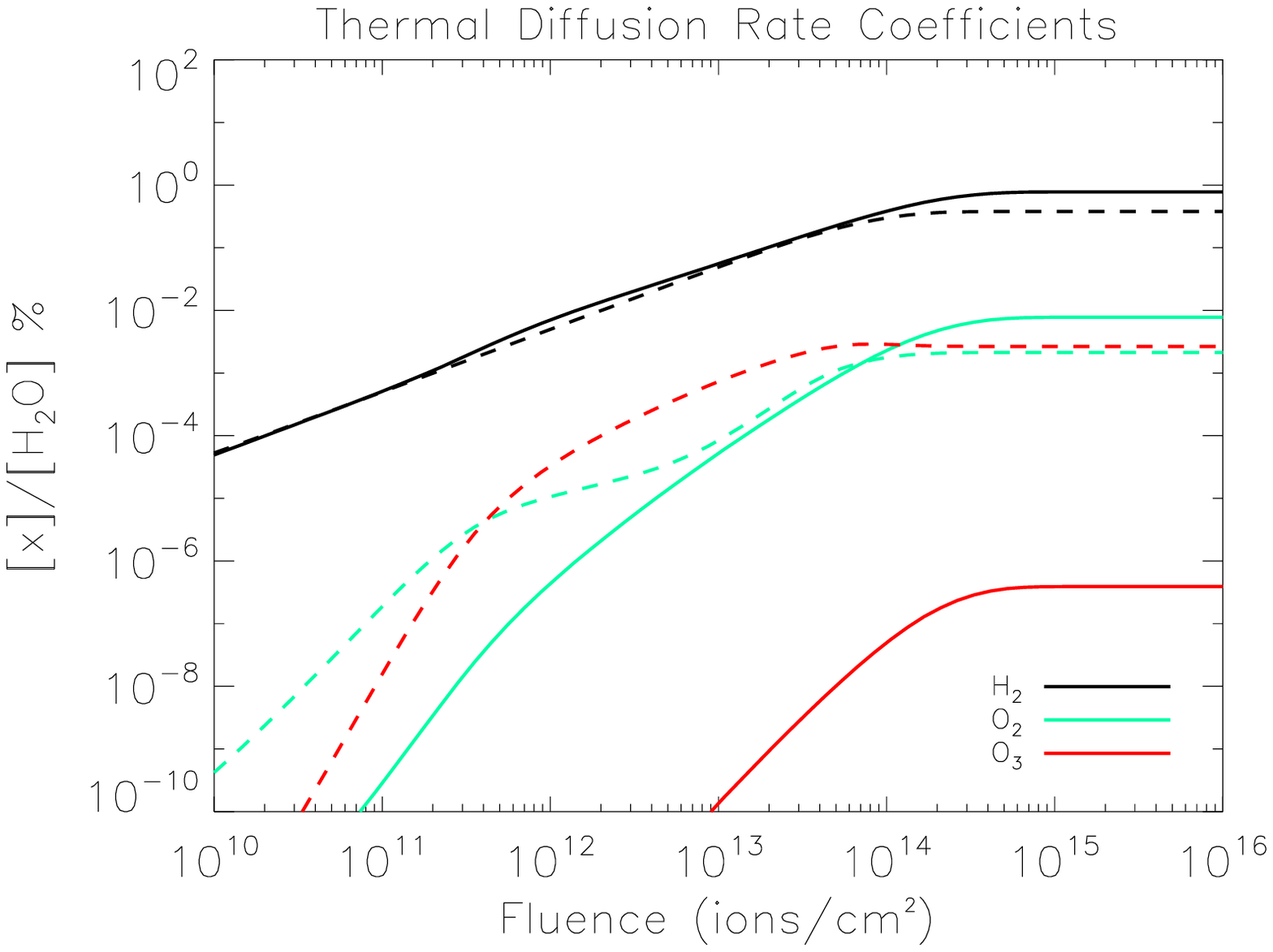}{0.45\textwidth}{(b)}
          \fig{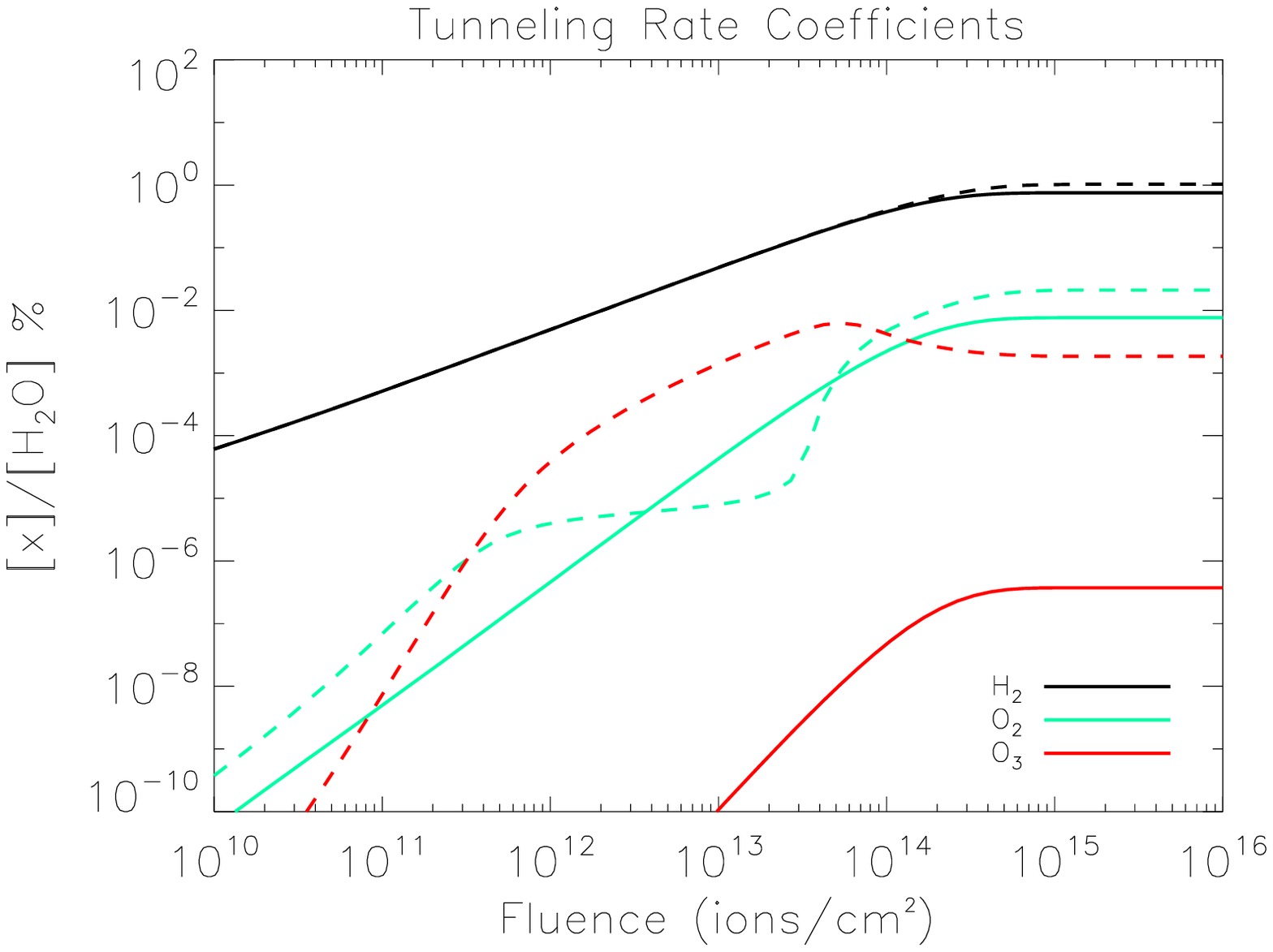}{0.45\textwidth}{(c)}
          }
\caption{
Calculated abundances of \ce{H2}, \ce{O2}, and \ce{O3} at 16 K (solid line) and 77 K (dashed line) from simulations of a pure \ce{H2O} ice bombarded by 200 keV \ce{H+} assuming (a) fast bulk reactions of radicals and atomic oxygen, (b) only thermal bulk diffusion, and (c) diffusion barrier tunneling for H, \ce{H2}, and O. 
\label{fig:others}}
\end{figure*}          


Shown in Fig. \ref{fig:others} are the abundances of \ce{H2}, \ce{O2}, and \ce{O3} versus fluence. In model (a), the abundance of \ce{O2} at 16 K is kept low because of destruction via \eqref{r6} to form \ce{HO2} but increases at 77 K due, in part, to more efficient formation via 

\begin{equation}
    \ce{HO2 + O3 -> O2 + O2 + OH}.
    \tag{R7}
    \label{r7}
\end{equation}

\noindent
which has a small barrier of 490 K \citet{burkholder_nasa_2014}. Similarly, the increase in molecular oxygen abundance at 77 K in model (c) is further driven by 

\begin{equation}
    \ce{OH + HO2 -> H2O + O2}
    \tag{R8}
    \label{r8}
\end{equation}

\noindent
where the abundances of OH and \ce{HO2} are enhanced relative to model (b) because of the effects of* quantum tunneling through diffusion barriers by H, \ce{H2}, and O - as shown in Fig. \ref{fig:radicals}. The decreased abundance of these radicals at 77 K in model (b), combined with destruction with atomic oxygen, leads to the drop in [\ce{O2}] in Fig. \ref{fig:others}b.

Unfortunately, comparison of our \ce{O2} results with experimental data, as with \ce{H2O2}, is complicated by the fact that homonuclear diatomic molecules, lacking permanent dipoles, are IR inactive. Thus, their abundances cannot be measured using standard techniques, such as Fourier Transform Infrared Spectroscoppy (FTIR). Nevertheless, it is well known that \ce{H2} and \ce{O2} form during water ice radiolysis based on  analysis of both sputtering products as well as post-irradiation temperature-programmed desorption (TPD) of the ice via mass spectrometry \citep{johnson_photolysis_1997,teolis_water_2017}. In principle, though, such measurements should be possible using Raman techniques \citep{rothard_modification_2017} and would be of great value, in part, by enabling us to further refine both our radiochemical yields and chemical network. Interest in constraining \ce{O2} abundances in irradiated water was recently renewed following its detection in the coma of comet 67P/C-G by \citet{bieler_abundant_2015}. As can be seen in Fig. \ref{fig:others}, the maximum abundance of \ce{O2} with respect to water achieved here is $\sim0.1$\% in model (a) at 77 K, a value which increased only negligibly at still higher temperatures. Thus, our models predict that the radiolysis of pure \ce{H2O} ice is not the dominant mechanism behind the $\sim3.8$
\% \ce{O2} abundances relative to water measured by Rosetta \citep{bieler_abundant_2015}. In that study, moreover, no evidence for ozone was found, though an upper limit of $1\times10^{-4}$ \% relative to water was established. Interestingly, only in model (a) are the ozone abundances predicted to remain below this limit, even at 77 K.

\section{Conclusions \& Outlook} \label{sec:conclusions}

In this work, we have simulated the bombardment of pure \ce{O2} and \ce{H2O} ices by energetic protons using a general rate-equation-based astrochemical code, modified to include radiation-chemical processes using the SH method. These models were carried out with the \texttt{MONACO} program \citep{vasyunin_formation_2017}, and a network consisting of the radiolysis processes listed in Table \ref{tab:gvalues} of Appendix \ref{sec:radiolysis} and the reactions noted in Appendix \ref{sec:network}. As illustrated in Figs. \ref{fig:o2ice} and \ref{fig:relative}, we were able to qualitatively reproduce both the abundance of \ce{O3} in pure \ce{O2} \citep{baragiola_solid-state_1999} and \ce{H2O2} in pure \ce{H2O} \citep{gomis_hydrogen_2004-1} utilizing the radiochemical processes given in Table \ref{tab:gvalues}. Thus validated, these processes, along with the reactions given in Table \ref{tab:h2onetwork}, can be added to existing chemical networks in order to better account for physicochemical effects caused by cosmic ray bombardment of dust grain ice-mantles.

Moreover, by simulating well-constrained experiments rather than interstellar environments we have been afforded a unique opportunity to compare the accuracy and physical realism of several approaches to modeling bulk chemistry over a variety of temperatures relevant to the ISM. As reported here, we have found that the standard approaches to bulk chemistry based on thermal diffusion or quantum tunneling through diffusion barriers did more poorly at reproducing the experimental data - particularly at low temperatures - than our model in which radicals and atomic oxygen were assumed to react quickly with neighboring species in the ice. This finding is in agreement with recent experiments by \citet{ghesquiere_reactivity_2018}, who found no evidence for true bulk diffusion. 

Regrettably, despite the large body of work in laboratory astrophysics on the irradiation of interstellar ice-analogues, it has not been possible, until recently, to incorporate many of the results of these experiments into astrochemical codes \citep{shingledecker_cosmic-ray-driven_2018}. However, our work presented here shows that not only can such models simulate radiation-chemical reactions, they might even be fruitfully used as a replacement for the simpler kinetic models sometimes employed (e.g. \citet{gomis_hydrogen_2004,baragiola_solid-state_1999}) in understanding and analyzing experimental data.

In summary, this study represents an attempt to shrink the existing gap between experiments and models, an increasingly urgent task in light of the upcoming launch of JWST. However, there is ample opportunity for even further refinements to our approach by, for example, considering the effects of the implantation and subsequent reactions of the bombarding \ce{H+} ions, of ice heating along the particle track, or of the effects caused by the nuclear/elastic component of the stopping, which begins to dominate over the electronic/inelastic component considered here at lower particle energies \citep{spinks_introduction_1990}. In addition to the synthesis of molecules, charged particle bombardment is well known to drive the non-thermal desorption of even large molecules such as benzene \citep{thrower_highly_2011,marchione_efficient_2016}. From experiments it is known that, particularly in water ice, excitons migrating to the surface represent one such mechanism that can stimulate this desorption. Given lingering questions about how molecules formed in dust-grain ice mantles are introduced into the surrounding gas in cold environments, future improvements to our approach in this area are warranted. Finally, experiments in which the abundances of multiple species are followed during irradiation would further advance our knowledge of radiation-chemical processes in ices and help to reduce uncertainties in future modeling research. In particular, Raman spectroscopic analysis - where even the behavior of IR-inactive species like \ce{O2} could be monitored - represents a powerful, yet perhaps underutilised, technique that should be considered in future studies.

\acknowledgments

CNS gratefully acknowledges the support of the Alexander von Humboldt Foundation. EH acknowledges the support of the National Science Foundation through grant AST-1514844. AV acknowledges the support of the Russian Science Foundation through grant 18-12-00351.

\software{MONACO \citep{vasyunin_formation_2017}}

\FloatBarrier

\appendix

\setcounter{table}{0}
\renewcommand{\thetable}{A\arabic{table}} 

\section{Radiolysis Processes \label{sec:radiolysis}}

The radiolysis processes and radiochemical yields used here are listed in Table \ref{tab:gvalues}.

\startlongtable
\begin{deluxetable}{lllc}
\tablecaption{Radiolysis reactions, branching fractions, and yields ( G-values) used in this work \label{tab:gvalues}}
\tablehead{
\colhead{Process} & \colhead{$f_\mathrm{br}$} & \colhead{$G$-value\tablenotemark{a}} & \colhead{Type}
}
\decimals
\startdata
\multicolumn{4}{c}{\ce{H2O}} \\
$\mathrm{H_2O \leadsto OH^* + H^*}$ & 0.9 & $2.57\times10^{-1}$ & I \\
$\mathrm{H_2O \leadsto O^* + H_2^*}$ & 0.1 & $2.57\times10^{-1}$ & I \\
$\mathrm{H_2O \leadsto OH + H}$ & 0.9 & $1.58\times10^{-1}$ & II \\
$\mathrm{H_2O \leadsto O + H_2}$ & 0.1 & $1.58\times10^{-1}$ & II \\
$\mathrm{H_2O \leadsto H_2O^*}$ & 1.0 & $1.58\times10^{-1}$ & III \\ \hline
\multicolumn{4}{c}{\ce{O2}} \\
$\mathrm{O_2 \leadsto O^* + O^*}$ & 1.0 & $6.81\times10^0$ & I \\
$\mathrm{O_2 \leadsto O + O}$ & 1.0 & $2.91\times10^0$ & II \\
$\mathrm{O_2 \leadsto O_2^*}$ & 1.0 & $2.91\times10^0$ & III \\ \hline
\multicolumn{4}{c}{\ce{O3}} \\
$\mathrm{O_3 \leadsto O_2^* + O^*}$ & 1.0 & $5.57\times10^1$ & I \\
$\mathrm{O_3 \leadsto O_2 + O}$ & 1.0 & $1.80\times10^1$ & II \\
$\mathrm{O_3 \leadsto O_3^*}$ & 1.0 & $1.80\times10^1$ & III \\ \hline
\multicolumn{4}{c}{\ce{HO2}} \\
$\mathrm{HO_2 \leadsto O^* + OH^*}$ & 0.5 & $3.70\times10^0$ & I \\
$\mathrm{HO_2 \leadsto H^* + O_2^*}$ & 0.5 & $3.70\times10^0$ & I \\
$\mathrm{HO_2 \leadsto O + OH}$ & 0.5 & $3.71\times10^0$ & II \\
$\mathrm{HO_2 \leadsto H + O_2}$ & 0.5 & $3.71\times10^0$ & II \\
$\mathrm{HO_2 \leadsto HO_2^*}$ & 1.0 & $3.71\times10^0$ & III \\ \hline
\multicolumn{4}{c}{\ce{H2O2}} \\
$\mathrm{H_2O_2 \leadsto OH^* + OH^*}$ & 0.5 & $5.10\times10^1$ & I \\
$\mathrm{H_2O_2 \leadsto H^* + HO_2^*}$ & 0.5 & $5.10\times10^1$ & I \\
$\mathrm{H_2O_2 \leadsto OH + OH}$ & 0.5 & $4.13\times10^1$ & II \\
$\mathrm{H_2O_2 \leadsto H + HO_2}$ & 0.5 & $4.13\times10^1$ & II \\
$\mathrm{H_2O_2 \leadsto H_2O_2^*}$ & 1.0 & $4.13\times10^1$ & III \\ \hline
\multicolumn{4}{c}{\ce{O}} \\
$\mathrm{O \leadsto O^*}$ & 1.0 & $3.70\times10^0$ & I \\
$\mathrm{O \leadsto O^*}$ & 1.0 & $1.93\times10^0$ & III \\ \hline
\multicolumn{4}{c}{\ce{H2}} \\
$\mathrm{H_2 \leadsto H^* + H^*}$ & 1.0 & $3.70\times10^1$ & I \\
$\mathrm{H_2 \leadsto H + H}$ & 1.0 & $1.02\times10^1$ & II \\
$\mathrm{H_2 \leadsto H_2^*}$ & 1.0 & $1.02\times10^1$ & III \\ \hline
\multicolumn{4}{c}{\ce{OH}} \\
$\mathrm{OH \leadsto O^* + H^*}$ & 1.0 & $3.70\times10^0$ & I \\
$\mathrm{OH \leadsto O + H}$ & 1.0 & $5.66\times10^0$ & II \\
$\mathrm{OH \leadsto OH^*}$ & 1.0 & $5.66\times10^0$ & III \\
\enddata
\tablenotetext{a}{Given in molecules/100 eV}
\tablecomments{Following \citet{shingledecker_cosmic-ray-driven_2018} we use Type I to indicate process \eqref{p2}, Type II process \eqref{p3}, and Type III process \eqref{p4}.}
\end{deluxetable}

\section{Chemical Network \label{sec:network}}

\setcounter{table}{0}
\renewcommand{\thetable}{B\arabic{table}}

The chemical network used in this work consists of the reactions listed in Table \ref{tab:h2onetwork}, as well as the neutral-neutral reactions in Table 7 of \citet{shingledecker_new_2017}.  

\startlongtable
\begin{deluxetable}{lrrr}
\tablecaption{Solid-phase reactions used in water simulations \label{tab:h2onetwork}}
\tablehead{
\colhead{Reaction} & \colhead{$E_\mathrm{A}$ (K)} & \colhead{$f_\mathrm{br}$} & \colhead{Source}
}
\decimals
\startdata
\ce{H + H -> H2} & 0 & 1.0 & \citet{wakelam_2014_2015} \\
\ce{H + H2 -> H2 + H} & 1\,900 & 1.0 & \citet{cohen_cumulative_1992} \\
\ce{H + O -> OH} & 0 & 1.0 & \citet{atkinson_evaluated_2004} \\
\ce{H + O2 -> HO2} & 0 & 1.0 & JL \\
\ce{H + O3 -> O2 + OH} & 450 & 1.0 & \citet{tielens_model_1982} \\
\ce{H + OH -> H2O} & 0 & 1.0 & \citet{atkinson_evaluated_2004} \\
\ce{H + H2O -> OH + H2} & 9\,700 & 1.0 & \citet{baulch_evaluated_1992} \\
\ce{H + HO2 -> O + H2O} & 0 & 0.0194 & \citet{atkinson_evaluated_2004} \\
\ce{H + HO2 -> O2 + H2} & 0 & 0.0857 & \citet{atkinson_evaluated_2004} \\
\ce{H + HO2 -> H2O2} & 0 & 0.0894 & \citet{atkinson_evaluated_2004} \\
\ce{H + H2O2 -> H2O + OH} & 1\,400 & 0.999 & \citet{baulch_evaluated_1992} \\ 
\ce{H + H2O2 -> HO2 + H2} & 1\,900 & 0.0006 & \citet{baulch_evaluated_1992} \\
\ce{H2 + O -> H2O} & 9\,700 &  1.0 & \citet{javoy_elementary_2003}  \\
\ce{H2 + O2 -> HO2 + H} & 28\,000 & 1.0 & \citet{karkach_ab_1999} \\
\ce{H2 + O3 -> OH + O2} & 10\,000 & 1.0 & See text \\
\ce{H2 + OH -> H2O + H} & 1\,800 & 1.0 & \citet{burkholder_nasa_2014} \\
\ce{H2 + H2O -> H2O + H}  & 10\,000 & 1.0 & See text \\
\ce{H2 + HO2 -> H2O2 + H} & 13\,000 & 1.0 & \citet{tsang_chemical_1986} \\
\ce{H2 + H2O2 -> H2 + OH + OH} & 10\,000 & 1.0 & See text \\
\ce{O + O2 -> O3} & 180 & 1.0 & \citet{benderskii_diffusionlimited_1996} \\
\ce{O + OH -> HO2} & 0 & 1.0 & \citet{atkinson_evaluated_2004} \\
\ce{O + H2O -> H2O2} & 8\,800 & 1.0 & \citet{karkach_ab_1999} \\
\ce{O + HO2 -> O2 + OH} & 0 & 1.0 & M84 \\
\ce{O + H2O2 -> OH + HO2} & 2\,000 & 1.0 & \citet{atkinson_evaluated_2004} \\
\ce{O2 + OH -> HO2 + O} & 25\,000 & 1.0 & \citet{srinivasan_reflected_2005} \\
\ce{O2 + H2O -> HO2 + OH} & 37\,000 & 1.0 & \citet{mayer_activation_1968} \\
\ce{O2 + HO2 -> O3 + OH} & 10\,000 & 1.0 & See text \\
\ce{O2 + H2O2 -> HO2 + HO2} & 18\,000 & 1.0 & \citet{donaldson_bimolecular_2003} \\
\ce{O3 + OH -> HO2 + O2} & 940 & 1.0 & \citet{atkinson_evaluated_2004} \\
\ce{O3 + H2O -> H2O2 + O2} & 10\,000 & 1.0 & See text \\
\ce{O3 + HO2 -> O2 + O2 + OH} & 490 & 1.0 & \citet{burkholder_nasa_2014} \\
\ce{O3 + H2O2 -> O2 + O2 + H2O} & 10\,000 & 1.0 & See text \\
\ce{OH + OH -> H2O2} & 0 & 1.0 & \citet{atkinson_evaluated_2004} \\
\ce{OH + H2O -> H2O2 + H} & 40\,000 & 0.5 & \citet{lamberts_influence_2017} \\ 
\ce{OH + H2O -> H2O + OH} & 2\,100 & 0.5 & \citet{uchimaru_ab_2003} \\
\ce{OH + HO2 -> H2O + O2} & 0 & 1.0 & \citet{schwab_kinetics_1989} \\
\ce{OH + H2O2 -> H2O + HO2} & 760 & 1.0 & \citet{buszek_effects_2012} \\
\ce{H2O + HO2 -> H2O2 + OH} & 17\,000 & 1.0 & \citet{lloyd_evaluated_1974} \\
\ce{H2O + H2O2 -> OH + OH + H2O} & 10\,000 & 1.0 & See text \\
\ce{HO2 + HO2 -> H2O2 + O2} & 0 & 1.0 & See text \\
\ce{H2O2 + H2O2 -> H2O + OH + HO2} & 10\,000 & 1.0 & See text \\
\enddata
\tablecomments{In the absence of other experimental or theoretical data, we here assume $E_\mathrm{A} = 0$ K for radical-radical reactions and $10^4$ K for all others.}
\end{deluxetable}

\bibliography{bibliography}

\end{document}